\documentstyle[12pt]{article}

\begin{document}

\Large
\bf

\begin{center}
DECOUPLING OF TRANSLATIONAL AND ROTATIONAL MODES FOR A QUANTUM SOLITON
\end{center}

\begin{center}
A.Dubikovsky and K.Sveshnikov\footnote{\large E-mail: costa@bog.msu.su}
\end{center}

\normalsize
\it

\begin{center}
Department of Physics, Moscow State University, Moscow 119899, Russia
\end{center}

\leftskip 1 true cm
\rightskip 1 true cm
\baselineskip 10 pt

\vskip 0.5 true cm
\normalsize
\rm

A set of integral relations for rotational and translational
zero modes in the vicinity of the soliton solution are derived
from the particle-like properties of the latter and verified for
a number of models (solitons in 1+1-dimensions, skyrmeons in
2+1- and 3+1-dimensions, non-abelian monopoles). It is shown,
that by consistent quantization within the framework of
collective coordinates these relations ensure the correct
diagonal expressions for the kinetic and centrifugal terms in
the Hamiltonian in the lowest orders of the perturbation
expansion. The connection between these properties and virial
relations is also determined.

\leftskip 0 true cm
\rightskip 0 true cm
\baselineskip 11 pt
\vskip 0.5 true cm

\large
\rm

Motivated by success of the skyrmeon baryon models \cite{1},
there was  recently much interest in quantization of hedgehog-type
configurations including translational and rotational
degrees of freedom by means of suitable collective coordinates
\cite{2}. Generally, in this approach the corresponding quantum
Hamiltonian contains a full bilinear form in conjugated momenta
with nontrivial couplings between different collective variables
\cite{3,4,5}.  However, for a particle-like classical solution one
should expect additive diagonal contributions of kinetic and
centrifugal terms to the Hamiltonian, at least to the lowest
orders in the a ppropriate weak coupling expansion.  Actually,
it is a part of the general problem of decoupling upon
quantization of various soliton degrees of freedom, which takes
place for any type of field models with classical solutions. In
this paper we'll present a consistent general analysis of this
problem for translational and rotational variables, based on the
particle-like properties of the classical solution combined with
Lorentz covariance and virial relations.

Let us consider a field theory in $d+1$ space-time dimensions
described by the Lagrangean density ${\cal L}(\varphi)$,
which possesses a classical particle-like solution
$\varphi_{c}(x)$. For brevity the Lorentz and internal symmetry
group indices are suppressed. It is generally accepted, that if
in the rest frame $\varphi_{c}$ is static with finite and
localized energy density, then in quantum version of the theory
such configuration describes an extended particle.
Now we'll show, that there exists a set of
nontrivial integral relations, fulfilled by $\varphi_{c}(x)$, which
provide the validity of these assumptions.

For a given static solution $\varphi_{c}(\vec x)$ the moving one
is constructed via Lorentz boost , what results in the replacement
\equation
  x^i \to {\Lambda^{-1}}^i_{\nu}x^{\nu}
\label{1}
\endequation
in the arguments of $\varphi_{c}$, where $\Lambda^{\mu}_{\nu}$ is
the corresponding Lorentz matrix.  The momentum of the moving
solution is
\equation
  P^{\mu}=\int T^{\mu 0}
   \big( \varphi_{c} ( \vec x,x^{0}) \big) \; d \vec x,
\label{2}
\endequation
where $T^{\mu\nu}(\varphi_{c})$ is the energy--momentum tensor.
Transforming the r.h.s. in (\ref{2}) to the rest frame, one gets
\equation
  P^{\mu}= \Lambda^{\mu}_{\mu'}\Lambda^{0}_{\nu'} \int
  T^{\mu'\nu'} \big( \varphi_{c} (\vec \xi) \big) \; J \; d \vec \xi,
\label{3}
\endequation
where $J={{\Lambda}^{0}_{0}}^{-1}$ is the Jacobian of transition
from $d \vec x$ to the rest frame spatial variable. On the other hand,
the l.h.s. of (\ref{2}) should be the momentum of a particle with
the mass $M$, that is
\equation
  P^{\mu}=\Lambda^{\mu}_{0}M.
\label{4}
\endequation
Then it follows from eqs.(\ref{3}) and (\ref{4}) that
\equation
  \int T^{\mu \nu}\big( \varphi_{c}(\vec \xi)\big) \; d \vec  \xi
         = \delta^{\mu}_{0}  \delta^{\nu}_{0}M.
\label{5}
\endequation
For $\nu=0$ we get $P^{0}=M,\; P^{i}=0,$ just that we should expect
for a static solution.  However, for $\mu=i, \; \nu=j$ we obtain
\equation
  \int { \partial {\cal L} (\varphi_c)  \over \partial \partial^{j}
    \varphi_{c}(\vec \xi)} \; \partial_{i} \varphi_{c}(\vec \xi) \;
       d \vec \xi= M  \delta_{ij}.
\label{6}
\endequation
(Henceforth the derivative $\partial_{i} $ acts on the indicated
argument of the function.) So we get the first set of conditions
(\ref{6}), which holds for a particle-like classical configuration
$\varphi_{c}(\vec \xi)$ in the rest frame.

Now let us consider  the orbital part of the 4-rotation tensor
(without the spin term)
\equation
  L^{\mu \nu}=\int \left( x^{\nu}T^{\mu 0}
   \big( \varphi_{c} (\vec x,x^{0}) \big) -x^{\mu}T^{\nu 0}
    \big( \varphi_{c} (\vec x,x^{0}) \big) \right) d \vec x.
\label{7}
\endequation
Returning to the rest frame, we can write similarly
\equation
 L^{\mu\nu}=\Lambda^{\mu}_{\mu'}
		 \Lambda^{\nu}_{\nu'}
                 \Lambda^{0}_{\sigma}
  \int \left(
      \xi^{\nu'}T^{\mu'\sigma} \big( \varphi_{c}(\vec \xi) \big)
     -\xi^{\mu'}T^{\nu'\sigma} \big( \varphi_{c}(\vec \xi) \big)
       \right) \; J \; d \vec \xi.
\label{8}
\endequation
On the other hand, for a static particle-like solution in the
rest frame one obviously has $L^{ij}=0$, while $L^{0i}$
coincide with the center-of-mass coordinates and therefore can be
made vanish by a spatial translation.  But since $\L^{\mu \nu}$
is a tensor, then it must vanish in any other Lorentz system.
Then it follows from eq.(\ref{8}) that
\equation
  0= \int \left( \xi^{\nu}T^{\mu\sigma} \big( \varphi_{c}(\vec \xi) \big)
    -\xi^{\mu}T^{\nu\sigma} \big( \varphi_{c}(\vec \xi) \big) \right)  \;
           d \vec \xi .
\label{9}
\endequation
For $\sigma=0$ these relations mean, that in  the rest frame
$L^{\mu \nu}=0$, just that we expected to have. When
$\mu=0$ or $\nu=0$ but $\sigma \not = 0$, then due to eq.(\ref{5})
one obtains from (\ref{9}) that $\int \xi^i T^{0j}( \varphi_{c})
d \xi =0$, what gives an identity provided by symmetry of
$T^{\mu \nu}$.  The latter statement is valid even for theories
with Chern--Simons terms, since these terms do not contribute to
$T^{\mu \nu}$ \cite{6}.  However, for $\mu=i$, $\nu=j$, $\sigma=k$
we get from (\ref{9}) the following relations (for definiteness,
we take $d=3$)
\equation
  \int \varepsilon_{lij} \; \xi_{i} \;
\partial_{j} \varphi_{c}(\vec \xi) {\partial {\cal L} (\varphi_c) \over
\partial \partial^{k} \varphi_{c}(\vec \xi)} \; d \vec \xi = \int
\varepsilon_{lik} \; \xi_{i} \; {\cal L} (\varphi_c)  \; d \vec \xi =0,
\label{10}
\endequation
since $L^{0k}$ vanish in the rest frame by assumption. This is
the second set of relations on $\varphi_{c}(\vec \xi)$,
following from the Lorentz covariance and particle-likeness
of the classical solution.

So each particle-like solution should be subject of conditions
(\ref{6}) and (\ref{10}). It should be noted, that the  relation
(\ref{4}) for $\mu=0$ reproduces nothing else but the relativistic
mass-energy relation.  For the moving $\varphi^4$-kink solution
this relation has been explicitly verified in \cite{5}, and for
the moving skyrmeon --- in (\ref{7}) by direct calculations.
However, the eqs.(\ref{6}) are more general and, moreover,
the eqs.(\ref{10}) also take place.  Note also,
that these relations, being consistent with the field equations
and conservation laws, do not be the direct consequences of the
latters, and should be considered separately.

As a direct result of these relations we get the
orthogonality of  the zero--frequency eigenfunctions in the
neighborhood of the classical particle-like solution cite{8}.
Let us discuss the theory of a nonlinear scalar field in $3$ spatial
dimensions, described by the Lagrangean density
\equation
  {\cal L} = {1 \over 2}{(\partial_{\mu} \varphi)}^{2}- U(\varphi),
\label{11}
\endequation
which possesses a  classical static solution
$\varphi_{c}(x)=u(\vec x).$ According to the virial theorem
such solutions are  unstable in more then one spatial dimension,
but for our purposes this is not so important compared to
simplicity of presentation.  In the general case the non-spherical
configuration $u(\vec x)$ yields 6 zero-frequency modes ---
three translational ones
\equation
  \psi_{i}(\vec x)=\partial_{i}u(\vec x),
\label{12}
\endequation
and three rotational
\equation
  f_{i}=\varepsilon_{ijk}x_{j}\partial_{k}u(\vec x).
\label{13}
\endequation
Then from eqs.(\ref{6}) and (\ref{10}) one immediately obtains
\eqnarray
  \int d \vec \xi \; \psi_{i}(\vec \xi) \; \psi_{j}(\vec \xi)
      & = & M\delta_{ij},
\label{14}
\\
   \int d \vec \xi \; f_{i}(\vec \xi) \; \psi_{j}(\vec \xi)
      & = & 0.
\label{15}
\endeqnarray
Further, by spatial rotations one can always achieve that
\equation
  \int d \vec \xi \; f_{i}(\vec \xi) \; f_{j}(\vec \xi)=
          \Omega_{ij}=\Omega_{i}\delta_{ij},
\label{16}
\endequation
where $\Omega_{i}$ are the moments of inertia of the classical
configuration.  Obviously, the relations (\ref{14}) and (\ref{15})
remain unchanged. As a result, the normalized set of translational
and rotational zero-modes can always be written as
$\{ \psi_{i}(\vec x)/ \sqrt {M} \ , \  f_{i}(\vec x)/\sqrt{\Omega_{i}}
\ \}.$

So the  particle-likeness of the classical solution results
in the diagonality of the zero-frequency scalar product matrix.
This diagonality plays an essential role in the procedure of
quantization in the vicinity of a classical soliton solution
by means of collective coordinates \cite{3,4,5}. Following the
conventional procedure \cite{4}, let us consider the field
$\varphi(\vec x)$ in the Schroedinger picture in the vicinity
of the solution $u(\vec x).$ The substitution, introducing
translational and rotational collective coordinates, reads \cite{8}
\equation
  \varphi(\vec x) = u\left( R^{-1}(\vec c)(\vec x-\vec q)\right)
            +\Phi\left( R^{-1}(\vec c)(\vec x-\vec q)\right),
\label{17}
\endequation
where $\Phi$ is the meson field, $R(\vec c)$ is the rotation matrix,
$\vec q$ and $\vec c$ are the translational and rotational
collective coordinates correspondingly.

For our purposes the parametrization of the rotation group
by means of the vector--parameter \cite{9} is the most convenient.
In this parametrization the rotation matrix $R(\vec c)$ is taken
in the form
\equation
   R(\vec c)=1+2 \; {c^{\times}+{c^{\times}}^{2} \over 1+c^{2}}
    ={1-c^{2}+2c^{\times}+2c\cdot c \over 1+c^{2}},
\label{18}
\endequation
where $c^{\times}_{ab}=\varepsilon_{adb} c_{d}$,
${(c\cdot c)}_{ab}=c_{a}c_{b}$, $c^{2}=\vec c \; \vec c.$
The composition law for vector--parameters, corresponding to product
of rotations $R(\vec a) \; R(\vec b)=R(\vec c),$
is given by
\equation
    \vec c=\langle \vec a, \; \vec b \rangle={\vec a+\vec b+\vec a \times
               \vec b \over 1-\vec a \; \vec b}.
\label{19}
\endequation
The generators of infinitesimal rotations are
\equation
    \vec S=-{i \over 2} \left( 1+c\cdot c +c^{\times} \right)
          {\partial \over  \partial \vec c} \; ,
\label{20}
\endequation
while the finite rotations $U(\vec a)$, defined so that $U^{+}(\vec a)
\; \vec b \; U(\vec a)=\langle \vec a, \; \vec b \rangle$, take the form
\equation
   U(\vec a)= \exp\left\{-2i\vec a\vec S\right\}.
\label{21}
\endequation

Returning to the decomposition (\ref{17}), one finds that the
total momentum of the field is now represented as
\equation
   \vec P=-i{\partial \over \partial \vec q} \; ,
\label{22}
\endequation
and the total angular momentum is equal to
\equation
   \vec J=\vec L+\vec S,
\label{23}
\endequation
where $\vec L=\vec q \times \vec P$ is the orbital angular
momentum, and the spin $\vec S$ is defined by relation (\ref{20}).

In order to keep the number of degrees of freedom we impose on
the field $\Phi(\vec y)$ 6 subsidiary conditions, which in the
theory of a weak coupling are usually taken as linear combinations
\equation
  \int d \vec y \; N^{(\alpha)}(\vec y) \; \Phi(\vec y)=0,
       \quad \alpha=1,\ldots,6.
\label{24}
\endequation
The set  $\{ N^{(\alpha)}(\vec y) \}$ should ensure the condition
of orthogonality of the meson field $\Phi(\vec y)$ to
zero-frequency modes and is chosen in the following way.
Let us denote $M^{(\alpha)}(\vec y) = \{ \psi_{i}(\vec y), \;
f_{i}(\vec y) \}.$ In the general case \cite{4} $N^{(\alpha)}(\vec y)$
are given by linear combinations of $M^{(\beta)}(\vec y)$ subject of
relations
\equation
  \int d \vec y \; N^{(\alpha)}(\vec y) \; M^{(\beta)}(\vec y)=
   \delta_{\alpha \beta}.
\label{25}
\endequation
In our case the system of zero-frequency modes is orthogonal,
so one immediately gets
\equation
  N^{(\alpha)}(\vec y)=\{\psi_{i}(\vec y)/M, \; f_{i}(\vec y)/\Omega_{i}\}.
\label{26}
\endequation
It is the relation (\ref{26}), that ensures the additive form  of
the collective  coordinate part of the Hamiltonian within the
weak coupling expansion in powers of the meson fields.  Let us
consider the condition (\ref{24}) as relation, defining $\vec q$ and
$\vec c$ as functionals of $ \varphi(\vec x)$. A straightforward
calculation gives
\eqnarray
  &  & N^{(\alpha)} \left( R^{-1}(\vec c)(\vec x- \vec q) \right)+
\nonumber
\\
     &  &  +R_{ji}{\partial R_{jk} \over \partial c_{l}}
          {\partial c_{l} \over \partial \varphi(\vec x)}
            \int d \vec y \; N^{(\alpha)}(\vec y) \;
     y_{k} \; \partial_{i} \left( u(\vec y)+\Phi(\vec y) \right)+
\label{27}
\\
     &  & +R_{ji}{\partial q_{j} \over \partial \varphi(\vec x)} \int
   d \vec y \; N^{(\alpha)}(\vec y) \; \partial_{i} \left( u(\vec y)
            +\Phi(\vec y) \right)=0.
\nonumber
\endeqnarray
By means of eqs.(\ref{18})--(\ref{21}) one can easily verify that
\eqnarray
   R_{ji} {\partial R_{jk} \over \partial c_l} & = & -{2 \over 1+c^2 } \
        \varepsilon_{ikm} (1-c^\times)_{ml},
\nonumber
\\ & \label{28} & \\
     (1-c^\times)^{-1} & = & {1+c \cdot c+c^\times \over 1+c^2}.
\nonumber
\endeqnarray
The simplicity of these relations demonstrates the convenience
of vector parametrization (\ref{18})--(\ref{21}) for such type of
problems \cite{8}.  Then from eqs.(\ref{27}), (\ref{28}) we
immediately get  the following lowest-order expressions for
$\partial \vec q / \partial \varphi(\vec x)$ and
$\partial \vec c / \partial \varphi(\vec x)$
\eqnarray
   -\vec \psi \left( R^{-1}(\vec c) (\vec x-\vec q) \right)
  & = & MR^{-1}{\partial \vec q  \over \partial \varphi(\vec x)},
\nonumber
\\ & \label{29} & \\
   -\vec f \left( R^{-1}(\vec c)(\vec x-\vec q)  \right)
  & = & 2 \; \Omega \; {1-c^{\times} \over 1+c^{2}}
            {\partial \vec c \over \partial \varphi(\vec x)}.
\nonumber
\endeqnarray
Calculating the conjugate momentum $\pi(\vec x)=-i\delta /
\delta \varphi(\vec x)$ as a composite derivative
\eqnarray
  \pi(\vec x) & = & -i{\delta \over \delta \varphi(\vec x)}
     = {\partial \vec c \over \partial \varphi (\vec x)}
        \left( -i{\partial \over \partial \vec c } \right)+
\nonumber
\\
 & + & {\partial \vec q \over \partial \varphi(\vec x)} \left(
   -i{\partial \over \partial \vec q} \right) +\int d \vec y \;
   {\delta\Phi(\vec y) \over \delta\varphi(\vec x)} \left( -i{\delta \over
      \delta\Phi(\vec y)} \right),
\label{30}
\endeqnarray
and using the relations (\ref{29}), we obtain to the leading
order the following result
\eqnarray
  \pi(\vec x) & = & \Pi \left( R^{-1}(\vec c)(\vec
   x-\vec y) \right)
\nonumber
\\
    & - & {1 \over M} \vec \psi \left( R^{-1}(\vec c)
   (\vec x-\vec q) \right) \left( \vec K + \int d\vec y \;
   \big{(} (\vec \partial \Phi) \; \Pi \big{)}(\vec y) \right)
\label{31}
\\
   & - & \vec f \left( R^{-1}(\vec c)(\vec x-\vec q) \right) \Omega^{-1}
			 \left( \vec I+ \int d \vec y \; \big{(} ( [ \vec
   y \times \vec \partial ] \; \Phi ) \; \Pi \big{)} (\vec y) \right).
\nonumber
\endeqnarray
In eq.(\ref{31}) $\vec K =R^{-1}(\vec c)\vec P$ and
$\vec I=R^{-1}(\vec c) \vec S$ are the momentum and the spin of
the field, corresponding to the rotating frame, and the meson field
momentum $\Pi(\vec y)$ is defined as
\equation
  \Pi(\vec y)=\int d\vec z \; A(\vec
    z,\vec y) \left( -i {\delta \over \delta \Phi(\vec z)} \right),
\label{32}
\endequation
where $A(\vec x, \vec y)$ is the projection matrix on the
subspace, orthogonal to zero-frequency modes
\equation
   A(\vec x, \vec y) =
   \delta(\vec x-\vec y)- \sum \limits _{\alpha} M^{(\alpha)}(\vec x) \;
       N^{(\alpha)}(\vec y).
\label{33}
\endequation
Inserting eqs.(\ref{17}) and (\ref{31}) in the Hamiltonian
\equation
 H =\int d \vec x \; \left\{ {1 \over 2} \pi^{2}(\vec x) +{1
   \over 2}{(\vec \partial \; \varphi)}^{2} +U(\varphi) \right\},
\label{34}
\endequation
we obtain finally the following lowest-order expression
\eqnarray
    H & = & M+ \int d \vec y \; \left\{ {1 \over 2} \Pi^{2}+ {1
    \over 2} {(\vec \partial \Phi)}^{2}+ {1 \over 2} U''(u(\vec
    y))\Phi^{2} \right\}(\vec y)
\nonumber
\\
   & + & {1 \over 2M}{\left( \vec K +
    \int d \vec y \; (\vec \partial \Phi) \; \Pi(\vec y) \right)}^{2}
\label{35}
\\
  & + & {1 \over 2} \sum\limits_{i} { { \left( I_i+\int d\vec y
    \; ([\vec y\times \vec \partial]_i \;\Phi)\;\Pi(\vec y) \right) }^{2}
   \over \Omega_{i} } .
\nonumber
\endeqnarray

It is indeed such form of the Hamiltonian, that
provides to interpret  the resulting ground state as
non-relativistic particle with the mass $M$ and moments of
inertia $\Omega_{i}$.  So the correct form of the Hamiltonian
with additive kinetic and centrifugal terms, that means the
absence of correlations between translational and rotational
degrees of freedom, is ensured by the diagonality of
zero--frequency scalar product matrix (\ref{14})--(\ref{16}).
In turn, this is a direct consequence of relations (\ref{6}) and
(\ref{10}). Note also,  that this result will be actually valid
for any field model in the neighborhood of the suitable
soliton solution.

These general considerations can be easily illustrated by concrete
models. Firstly, we consider the theory of a scalar field
in 1+1-dimensions, described by  the Lagrangean density (\ref{11}).
In this case we have only one relation (\ref{6})
\equation
   \int dx \; { (\varphi'(x)) }^{2} = M,
\label{36}
\endequation
where the mass $M$ is given by
\equation
    M = \int dx \; {1 \over 2} { (\varphi'(x)) }^{2} +
         		   \int dx \; U(\varphi(x)).
\label{37}
\endequation
Performing the dilatation
$\varphi(x) \; \rightarrow \varphi(\lambda x)$
and demanding for the solution at $\lambda=1$, i.e.
${ \left( { dM(\lambda) \over d\lambda } \right) }_{\lambda=1}=0 $,
we find the well-known Hobart--Derrick virial relation [10]
\equation
   {1 \over 2} \int dx \; { (\varphi'(x)) }^{2} = \int dx \;
                        	     U(\varphi(x)),
\label{38}
\endequation
owing to which the "particle-likeness condition" (\ref{36})
is fulfilled  automatically.

In more spatial dimensions the situation with the model (\ref{11})
is more complicated. Namely, it is a trivial task to verify
by the same arguments, that for each $i$
\equation
  \int (\partial_{i} \varphi)^2 d \vec x=M.
\label{39}
\endequation
However, the orthogonality conditions between different
spatial derivatives, predicted by eqs.(\ref{14}) and (\ref{15}),
cannot be derived by such simple
considerations. So here the additional arguments, used by derivation of
relations (\ref{6}) and (\ref{10}), are crucial.

     In 2+1-dimensions, the solitons in $CP_{N}$-models are
interesting examples with such particle-like properties.
As it is well-known, for $N=1$ the $CP_{N}$-model is reduced to
$O(3)$-model \cite{11}, described by
\equation
   {\cal L} = {1 \over 2} \partial_{\mu} \varphi^{a}
                         \partial^{\mu} \varphi^{a}
\label{40}
\endequation
with subsidiary condition
\equation
    \varphi^{a} \varphi^{a} = 1.
\label{41}
\endequation
This theory is a planar analog of the Skyrme model \cite{12}.
The standard one-particle solution of the model is given by \cite{13}
\equation
  \varphi^{1}
      = \phi (r) \cos n \vartheta , \; \; \; \varphi^{2} = \phi (r) \sin n
      \vartheta , \; \; \; \varphi^{3} = { (1- \phi^{2} ) }^{1/2},
\label{42}
\endequation
where $r,\vartheta$ are polar coordinates and
\equation
   \phi (r) = { 4r^{n} \over r^{2n}+4 },
\label{43}
\endequation
and describes the ''baby-skyrmeon'' configuration with the topological
charge $Q=n$ and the mass $M=4 \pi Q$.  The direct insertion
of expression (\ref{42}) into conditions (\ref{6}) and (\ref{10})
yields
\equation
   \int \partial_{i}
  \varphi^{a} \partial_{j} \varphi^{a} d^{2}x = 4 \pi n \delta_{ij} = M
  \delta_{ij},
\label{44}
\endequation
and
\equation
  \int \varepsilon_{ij} x_{i} \partial_{j}
  \varphi^{a} \partial_{k} \varphi^{a} d^{2}x = 0,
\label{45}
\endequation
that means the particle-likeness of the solution (\ref{42})
in the way described above.

As a more nontrivial example, we consider the $SU(2)$-Skyrme model
in 3+1-dimensions \cite{1,14}, including the break--symmetry pion
mass term
\eqnarray
    {\cal L} &  = & -{ f^{2}_{\pi} \over 4 } \; tr \;
	  L^{2}_{\mu} + { 1 \over 32 g^{2} } \; tr \; { [ L_{\mu} L_{\nu}
	  ] }^{2}  +{ m^{2}_{\pi} \over 4 } \; tr \; ( U+U^{+}-2 )
\nonumber
\\
	    & = & \; {\cal L}^{(2)} + {\cal L}^{(4)} + {\cal L}_{B},
\label{46}
\endeqnarray
where, as usually, $L_{\mu}=U^{-1}\partial_{\mu}U$ is
the left chiral current and $U=\sigma+i\tau^{a}\pi^{a}$ is the
quaternion field. In the quaternion representation one has
\eqnarray
     {\cal L}^{(2)} & = &
           {f^{2}_{\pi} \over 2}{(\partial_{\mu}\sigma)}^{2}
          +{f^{2}_{\pi} \over 2}{(\partial_{\mu}\pi^{a})}^{2} ,
\nonumber
\\
      {\cal L}^{(4)} & = &
         -{1 \over 4g^{2}} \Big{(} {(\partial_{\mu}\pi^{a})}^{4}
	 -{(\partial_{\mu}\pi^{a}\partial_{\nu}\pi^{a})}^{2}
\label{47}
\\
       &  & \; \;
  +2\big{(}{(\partial_{\mu}\sigma)}^{2}{(\partial_{\nu}\pi^{a})}^{2}
            -\partial_{\mu}\sigma\partial_{\nu}\sigma
           \partial_{\mu}\pi^{a}\partial_{\nu}\pi^{a} \big{)}\Big{)},
\nonumber
\\
      {\cal L}_{B} &  =  &
                 m^{2}_{\pi}( \sigma - 1 ).
\nonumber
\endeqnarray
Supposing the conventional ''hedgehog'' {\it Ansatz}
\equation
   \sigma=\cos \phi (r), \;\;\;\;
	    \pi^{a}={r^{a} \over r}\sin \phi (r)
\label{48}
\endequation
we find for the mass of the skyrmeon
\eqnarray
        &  &  M=  M^{(2)}+M^{(4)}+M_{B},
\nonumber
\\
   &  &   M^{(2)}=  4 \pi f^{2}_{\pi} \int r^{2} dr \;{1 \over 2}
               \left( {\phi'}^{2}+{2 \over r^{2}}\sin^{2}\phi \right),
\label{49}
\\
 &  &   M^{(4)}=  {4\pi \over g^{2}}\int r^{2}dr\;{\sin^{2}\phi \over 2r^{2}}
	\left({\sin^{2}\phi \over r^{2}}+2{\phi'}^{2} \right),
\nonumber
\\
  &   &   M_{B}=  4 \pi m^{2}_{\pi} \int r^{2}dr \; 2\sin^{2}{\phi \over 2}.
\nonumber
\endeqnarray
A straightforward calculation gives
\equation
   \int {\partial {\cal L} \over \partial \partial^{j} u^{A}}
            \partial_{i} u^{A} d^{3}x =
      {2 \over 3}\delta_{ij}(M^{(2)}+2M^{(4)}),
\label{50}
\endequation
here $u^{A} = \{ \sigma, \; \pi^{a} \}.$  Inserting eqs.(\ref{49})
and (\ref{50}) into (\ref{6}), we obtain the first
"particle-likeness condition" for the skyrmeon
\equation
    M^{(2)}-M^{(4)}+3M_{B}=0.
\label{51}
\endequation
On the other hand, the scaling $u^{A}(\vec x) \rightarrow u^{A}(\lambda
\vec x)$ yields
\equation
   M(\lambda)={1 \over \lambda} M^{(2)}+ \lambda
			       M^{(4)}+ {1 \over \lambda^{3}} M_{B}.
\label{52}
\endequation
It is easy to verify, that the requirement of
${ \left( {dM(\lambda) \over d \lambda} \right)}_{\lambda=1}=0$
coincides with the eq.(\ref{51}) for the skyrmeon. So the
particle-likeness condition (\ref{51}) for the skyrmeon is
fulfilled due to virial relation.   Further, inserting the
substitution (\ref{48}) into eqs.(\ref{10}) we find in the same
way, that the second set of relations for the skyrmeon is
provided  by the symmetry properties. So we'll get upon
quantization, that the full bilinear form considered in \cite{2},
automatically simplifies up to a diagonal construction similar
to eq.(\ref{35}), and therefore the quantized skyrmeon describes
an extended non-relativistic particle.

     Finally, we consider the 't Hooft--Polyakov monopole
for the $SU(2)$-Yang--Mills--Higgs theory \cite{5,15}, described by the
Lagrangean
\equation
   {\cal L} = -{1 \over 4} { (F^{a}_{\mu \nu})
      }^{2} +{1 \over 2} { (D_{\mu} \phi) }^{2}-V(\phi),
\label{53}
\endequation
where
\eqnarray
    & & V(\phi)= {\lambda \over 4}{(\phi^{a} \phi^{a}- \eta^{2})}^{2},
\nonumber
\\
    & & F^{a}_{\mu \nu}=  \partial_{\mu} A^{a}_{\nu} -\partial_{\nu}
   A^{a}_{\mu} +g \varepsilon^{abc} A^{b}_{\mu} A^{c}_{\nu},
\label{54}
\\
    & & D_{\mu} \phi^{a}=  \partial_{\mu} \phi^{a} +g \varepsilon^{abc}
   		       A^{b}_{\mu} \phi^{c}.
\nonumber
\endeqnarray
The monopole solution is given by
\equation
   \phi^{a} = {1 \over g}{
     r^{a} \over r^{2} } H(r), \;\;\; A^{a}_{i}= {1 \over g}
    \varepsilon_{aij}{ r_{j} \over r^{2} } (1-K(r)), \;\;\; A^{a}_{0}=0.
\label{55}
\endequation
The mass of the monopole is equal to
\eqnarray
       M=4 \pi {\eta \over g} \int {d \xi \over \xi^{2}}
   &  &   \bigg{[} \xi^{2} {K'}^{2} + {1 \over 2} { (1-K^{2}) }^{2}
              +{1 \over 2} { (H- \xi H') }^{2}
\nonumber
\\
    \;\; &   & \qquad \qquad \qquad +  K^{2} H^{2}
	      +{\lambda \over 4g^{2}} { (H^{2}- \xi^{2} ) }^{2} \bigg{]},
\label{56}
\endeqnarray
where the integration variable $\xi$ is connected with the radial
coordinate $r$ via $\xi=g \eta r.$
The l.h.s of condition (\ref{6}) reads
\eqnarray
   \int {\partial {\cal L} \over \partial \partial^{j} u^{A}}
          \partial_{i} u^{A} d^{3}x  & = &
          4 \pi {\eta \over g} \cdot {2 \over 3} \delta_{ij}
      \int {d \xi \over \xi^{2}}
      \bigg{[} \xi^{2} {K'}^{2} +
\nonumber
\\
    +  2(1 & - & K)^{2} + {(1-K)}^{3}
	+ {1 \over 2}{(H-\xi H')}^{2} + H^{2}K \bigg{]},
\label{57}
\endeqnarray
here $u^{A}=\{A^{a}_{\mu}, \; \phi^{a} \}.$
Now we take into account of the virial considerations. Performing the
scaling $u^{A}(\vec x) \rightarrow u^{A}(\lambda \vec x)$, we find
\eqnarray
    M(\lambda) & = & {1 \over \lambda} M_{2} + {1 \over
    \lambda^{2}} M_{1} + {1 \over \lambda^{3}} M_{0} + {1 \over
		 \lambda^{3}} M_{V},
\nonumber
\\
    M_{2} & = & \int d^{3}x \; {1
	      \over 2} \left[ {(\partial_{\mu}A^{a}_{\nu})}^{2}
		    -\partial_{\mu}A^{a}_{\nu}\partial_{\nu}A^{a}_{\mu}
                    -{(\partial_{\mu}\phi^{a})}^{2} \right]=
\nonumber
\\
        &  = &
	      4 \pi {\eta \over g} \int {d \xi \over \xi^{2}}
              \left[ \xi^{2} {K'}^{2} + 2 {(1-K)}^{2}
	     +{1 \over 2} {(H-\xi H')}^{2} + H^{2} \right],
\nonumber
\\
     M_{1} & = &
              \int d^{3}x \; g
              \left[ \varepsilon^{abc}
                     \partial_{\mu}A^{a}_{\nu}A^{b}_{\mu}A^{c}_{\nu}
                    -\varepsilon^{abc}
                   \partial_{\mu}\phi^{a}A^{b}_{\mu}\phi^{c} \right]=
\label{58}
\\
         &  = &
              4 \pi {\eta \over g} \int {d \xi \over  \xi^{2}}
	      \left[ -2{(1-K)}^{3}-2H^{2}(1-K) \right],
\nonumber
\\
     M_{0} & = &
              \int d^{3}x \; {1 \over 2} g^{2}
              \left[ {1 \over 2} \left( {(A^{a}_{\mu})}^{4} -
                      {(A^{a}_{\mu}A^{a}_{\nu})}^{2} \right)
                    -\left( {(A^{a}_{\mu})}^{2}{(\phi^{b})}^{2}-
                      {(A^{a}_{\mu}\phi^{a})}^{2} \right) \right]=
\nonumber
\\
         & = &
              4 \pi {\eta \over g} \int {d \xi \over \xi^{2}}
	      \left[ {1 \over 2} {(1-K)}^{4} + H^{2}{(1-K)}^{2} \right],
\nonumber
\\
     M_{V} & = &
              \int d^{3}x \; V(\phi(x))
            =  4 \pi {\eta \over g} \int {d \xi \over \xi^{2}}
	      {\lambda \over 4g^{2}}{(H^{2}-\xi^{2})}^{2}.
\nonumber
\endeqnarray
The requirement
${ \left( {dM(\lambda) \over d \lambda} \right) }_{\lambda=1}=0$
yields
\eqnarray
      \int {d \xi \over \xi^{2}} \bigg{[}\xi^{2}{K'}^{2}
          + {1 \over 2} (1 & - &  K)^{2}  (3K^{2}+2K-1)
      + {1 \over 2} { (H- \xi H')}^{2}
                             \qquad \qquad
\label{59}
\\
 \;\;  & - &  2 H^{2} K + 3 H^{2} K^{2}
      + 3 {\lambda \over 4g^{2}} {(H^{2}-\xi^{2})}^{2} \bigg{]} = 0.
\nonumber
\endeqnarray
Inserting eqs.(\ref{57}) and (\ref{58}) into (\ref{6}), we see that the
particle-likeness condition (\ref{6}) coincides with
eq.(\ref{59}). The condition (\ref{10}) is fulfilled due to
symmetry properties of the expression (\ref{55}), just as in the  case of
skyrmeon.

     So we have  proved the validity of the "particle-likeness"
conditions (\ref{6}) and (\ref{10}) for the most important soliton
solutions. Note, that there is a close connection
between the condition (\ref{6}) and virial relations for the static
configuration by  homogeneous dilatations.  The relations (\ref{10})
are usually fulfilled on account of symmetry properties of
classical solutions. Since the Lorentz group is six-parametric
and there is only one parameter by homogeneous dilatations, then
it looks naturally, that the relations (\ref{6}) and (\ref{10}) should be
splitted in two parts:

    1) the l.h.s. of relations (\ref{6}) should be diagonal
\equation
   \int { \partial {\cal L} (\vec \xi) \over
	     \partial \partial^{j} \varphi_{c}(\vec \xi)} \;
      \partial_{i} \varphi_{c}(\vec \xi) \; d \vec \xi =
		 I \delta_{ij},
\label{60}
\endequation
and l.h.s. of eqs.(\ref{10}) should vanish
\equation
  \int \varepsilon_{lij} \; \xi_{i} \;
	    \partial_{j} \varphi_{c}(\vec \xi) \;
      {\partial {\cal L}(\vec \xi) \over
       \partial \partial^{k} \varphi_{c}(\vec \xi) } \; d\vec\xi=0,
\label{61}
\endequation
provided by symmetry properties of the static configuration;

    2) the value of $I$ coincides with the mass $M$ of the solution
\equation
    I=M
\label{62}
\endequation
due to the virial arguments.

The last statement is verified for all the interesting classical
solutions. However, the general proof of this connection is
not found yet.

To conclude let us  mention, that the present analysis can be
easily extended to internal degrees of freedom.  On the other
hand, the relations (\ref{6}) and (\ref{10}), being automatically
consistent with  exact solutions of equations of motion, can play an
essential role of additional constraints in approximate
calculations. For  example, they can be explored as a test for
various sample functions, used in describing the shape of the
skyrmeon \cite{1,16}. The results of this work will be reported
separately.

The authors are  indebted to Dr.A.Vlasov and P.Silaev for
interest and fruitful discussions.

\vskip 1.5 true cm

\normalsize
\rm

\end{document}